\documentclass[runningheads]{llncs}

\usepackage[round]{natbib}

\usepackage{times}
\usepackage{url}
\usepackage[utf8]{inputenc}
\usepackage[small]{caption}
\usepackage{graphicx}
\usepackage{amsmath}
\usepackage{booktabs}
\usepackage{algorithm}
\usepackage{algorithmic}

\usepackage{hyperref} 
\usepackage{color,soul} 
\usepackage{isabelle,isabellesym,isabelletags} 
\usepackage{amssymb} 
\isabellestyle{it}
\begin{document}

\title{Harnessing Higher-Order (Meta-)Logic to \\ Represent and Reason
  with Complex Ethical Theories\thanks{Supported by 
  VolkswagenStiftung, grant \textit{Consistent, Rational Arguments in Politics (CRAP)}.}}

\author{David Fuenmayor\inst{1}
	\and
	Christoph Benzm\"uller\inst{1,2} 
}
\authorrunning{D. Fuenmayor and C. Benzm\"uller}
\titlerunning{HOL for Reasoning with Ethical Theories}
%
\institute{Freie Universit\"at Berlin, Germany \and
	University of Luxembourg, Luxembourg}

\maketitle

\begin{abstract} 
	The computer-mechanization of an ambitious explicit ethical
        theory, Gewirth's Principle of Generic
	Consistency, is used to showcase an approach for representing and reasoning with ethical
	theories exhibiting complex logical features like alethic and deontic modalities, indexicals, higher-order quantification, among others. Harnessing the high expressive power of Church's type theory as a
	meta-logic to semantically embed a 
	combination of quantified non-classical logics, our
	work pushes existing boundaries in knowledge representation and
	reasoning. We demonstrate that intuitive encodings
	of complex ethical theories and their automation on
	the computer are no longer antipodes.
\end{abstract}

\section{Introduction} 
\label{sec:introduction}
\noindent 
Hybrid architectures for ethical autonomous agents that integrate both
bottom-up learning and top-down deliberation from upper principles are
receiving increased attention; cf.~\citet{dignum18:_special_issue,ijcai2017-655,scheutz2017case,malle2016integrating,DBLP:journals/ras/DennisFSW16,anderson2014geneth,Wallach} and the references therein. 
Irrespective of the preferred direction, it is becoming increasingly evident that adequate explicit representations of ethical knowledge are beneficial, if not mandatory, to obtain satisfactory solutions. 
Bottom-up approaches may benefit from expressive languages to \emph{explicitly} represent the learned ethical knowledge in an scrutable, communicable and transferable manner. Top-down approaches usually rely on expressive logic languages to enable an \emph{intuitive and accurate representation and reasoning} with ethical theories. Unfortunately, however, very few approaches are currently available that enable adequate and realistic, explicit formal encodings of non-trivialized ethical theories, and that at the same time support intuitive interactive-automated reasoning with them.

In this paper \emph{we demonstrate a methodology and implementation of
  such an ambitious ethical reasoning machinery}. Our approach is
based on classical higher-order logic (HOL), aka Church's type theory \citep{sep-type-theory-church}, which we exploit as a meta-logic to encode combinations of 
non-classical logics for normative reasoning as suited for a given application context.
The methodology and techniques we present, cf.~also \cite{J48}, can bring many benefits to the design of ethically-critical systems aiming at scrutability, verifiability, and the ability to provide justification for its decision-making. They are particularly relevant to the design of explicit ethical agents~\citep{Moor2009}. In particular, this area faces tough philosophical and practical challenges. No consensus is currently  in sight, if possible at all, concerning the choice of upper moral values and principles that constitute a generally agreed normative ethics for intelligent autonomous agents. For example, utilitarianism and deontology have both been critically discussed in this context.

\emph{We exemplarily study another relevant and ambitious theory in normative ethics: Alan Gewirth's ``Principle of Generic Consistency (PGC)" } \citep{GewirthRM,Beyleveld}, which has been proposed as an emendation of the \emph{Golden Rule}. Our aim is not to defend or assess Gewirth's work in comparison to other approaches.  We instead present a methodology and technique enabling the intuitive and accurate representation of ambitious ethical theories, and for this we take the PGC as a showcase and exemplarily assess its logical validity. Such an ambitious ethical theory has never before been
assessed on the computer at
such a level of detail  (i.e.~without trivializing it by abstraction).

Our method enables the reuse of modern interactive and automated higher-order theorem proving technology, and in this sense it establishes a \emph{relevant bridge between different research communities}.  On a practical level our work also addresses what we consider one of the biggest challenges in the area: to \emph{represent complex ethical theories in both a machine and human interpretable manner and to carry out complex reasoning in real-time with incomplete and inconsistent information}. And finally, as a side-effect, we have \emph{revealed and fixed some (minor) issues in Gewirth's PGC}.

Our choice of HOL  at the meta-level is motivated by the goal of flexibly combining expressive non-classical logics as required for the formal encoding of complex ethical theories. Current theories in normative and machine ethics are, quite understandably, formulated predominantly in natural language. While this supports human deliberation and agreement about what kind of moral beings we want future intelligent agents to be, it  also hampers their implementation in machines. Hence expressive formal languages are required, which enable flexible combinations of different types of non-classical logics. This is because ethical theories are usually challenged by complex linguistic expressions, including modalities (alethic, epistemic, temporal, etc.), counterfactual conditionals, generalized quantifiers, (un-)conditional obligations, among many others.

The meta-logical approach we exploit and demonstrate grounds on a technique known as \textit{(shallow) semantical embedding}. The approach will be addressed in \S\ref{sec:embedding}, 
where we present an extended embedding of a dyadic deontic logic (DDL)
by 
\cite{CJDDL} in HOL and \emph{combine, among others, conditional
  obligations with further modalities and quantifiers}. The combined
logic is immune to known paradoxes in deontic logic, in particular,
the so-called contrary-to-duty scenarios, in which a `secondary'
obligation must come into effect when a `primary' obligation is
violated (contradicted). Moreover, conditional (dyadic) obligations in
DDL are of a defeasible and paraconsistent nature and thus lend
themselves to normative reasoning with incomplete and inconsistent
information. In \S\ref{sec:representation} and \S\ref{sec:reasoning}
we will represent and formally assess Gewirth's PGC using this
expressive logic combination. We also demonstrate how our technique
has been utilized to reveal and fix some (minor) issues in Gewirth's
work. 
Related work and short summary are presented in
\S\ref{sec:Summary}, and a formally-verified, unabridged version of
our formal encoding of Gewirth's theory and argument is provided in
\cite{GewirthProofAFP}.

\section{Combining Expressive Logics in HOL} 
\label{sec:embedding}
\noindent{We utilize the
  \emph{shallow semantical embeddings} (SSE) approach to combining
  logics. 
  SSE exploits HOL as a
  meta-logic in order to embed the syntax and semantics of some
  target logics, thereby turning theorem proving systems
  for HOL into universal reasoning engines 
  \citep{J41}. Moreover, an approach drawing upon SSE has been
  proposed as the foundation for a flexible deontic logic reasoning
  infrastructure \citep{J48}. We thus 
  assess, in some sense, the promises of this framework at hand of a non-trivial,
  concrete example.

In the following, we present \textit{an extract} of the embedding of (extended) DDL
in HOL. Our work thereby extends previous work by
\cite{C71}: Besides adding higher-order
quantification, we also extend this embedding to a two-dimensional
semantics \citep{SEP2DSem} by additionally adding contextual
information; for this we use Kaplanian \textit{contexts of use}, cf.~\cite{Kaplan1979,Kaplan1989-KAPA}. The system platform used to implement
this ambitious logic combination is the Isabelle
proof assistant \citep{Isabelle}. In what follows, we are using Isabelle/HOL syntax to render axioms, theorems and definitions (providing the appropriate indications when
needed).\footnote{The formal content of this paper has been generated
	directly by Isabelle from our source files. A benefit is the
	prevention of typos.  As a side contribution we showcase the
	usability of modern proof assistants for the non-initiated in order to foster their application.} 

\subsection{Definition of Types}


The type $w$ corresponds to the original type for possible
worlds/situations in DDL, cf. \cite{C71}. We draw in this work upon David Kaplan's \textit{logic of indexicals/demonstratives} as originally presented in \cite{Kaplan1979}. In Kaplan's logical theory, entities of the aforementioned type $w$ would correspond to his so-called ``circumstances of evaluation''. Moreover, Kaplan introduces an additional dimension $c$, so-called ``contexts of use'', which allow for the modelling of particular context-dependent linguistic expressions, i.e. \textit{indexicals} (see section~\ref{sectionKaplan}). We additionally introduce some type aliases: $wo$ for intensions (also called ``contents'' or ``propositions'' in Kaplan's work), which are identified with their truth-sets i.e. the set of worlds at which the proposition is true, and $cwo$ (aliased $m$) for sentence meanings (also called ``characters'' in Kaplan's theory), which are modelled as functions from contexts to intensions. Moreover, a type $e$ for individuals is introduced to e.g.~enable quantification over individuals.}
\small \\
\isanewline
\isacommand{typedecl}\ w%
\ \ \isamarkupcmt{Type for possible worlds (Kaplan's ``circumstances of evaluation'')}
\isanewline
\isacommand{typedecl}\ c%
\ \ \isamarkupcmt{Type for Kaplan's ``contexts of use''}
\isanewline
\isacommand{typedecl}\ e%
\ \ \isamarkupcmt{Type for individuals}
\isanewline
\isacommand{type{\isacharunderscore}synonym}%
\ wo\ {\isacharequal}\ {\isachardoublequoteopen}w{\isasymRightarrow}bool{\isachardoublequoteclose}\ %
\isamarkupcmt{Type for contents/propositions%
}\isanewline
\isacommand{type{\isacharunderscore}synonym}%
\ cwo\ {\isacharequal}\ {\isachardoublequoteopen}c{\isasymRightarrow}wo{\isachardoublequoteclose}\ \ %
\isamarkupcmt{Type for sentence meanings (Kaplan's ``characters'')%
}\isanewline
\isacommand{type{\isacharunderscore}synonym}%
\ m\ {\isacharequal}\ {\isachardoublequoteopen}cwo{\isachardoublequoteclose}\ \ \ \ \ \ %
\isamarkupcmt{Type alias `m' for characters}
\normalsize

\subsection{Embedding of DDL Modal and Deontic Operators}
The semantics of DDL draws on Kripke semantics for its (normal) alethic modal operators and on a neighbourhood semantics\footnote{Neighbourhood semantics is a generalisation of Kripke semantics, developed independently by Dana Scott and Richard Montague. Whereas a Kripke frame features an accessibility relation $R: W{\rightarrow}2^W$ indicating which worlds are alternatives to (or, accessible from) others, a neighborhood frame $N: W{\rightarrow}2^{2^W}$ (or, as in our case, $N: 2^W{\rightarrow}2^{2^W}$) features a neighbourhood function assigning to each world (or set of worlds) a set of sets of worlds.} for its (non-normal) deontic operators. In order to embed those, we need to introduce the operators $av$ and $pv$ (which can be seen as accessibility relations between worlds), and $ob$ (denoting a neighborhood function operating on sets of worlds) at the meta-logical level. Several axioms, not shown here, adequately
constraint the interpretations of $av$, $pv$ and $ob$ (e.g. $av(w)$ is
always a subset of $pv(w)$). See~\cite{CJDDL} and~\cite{C71} for
further details.

The following Isabelle/HOL commands illustrate the way logical
operators in the target logic (enhanced DDL) can be defined as
metalogical predicates using lambda expressions of the appropriate
arity/type.
The two definitions below, introduced using Isabelle's keyword
``abbreviation", realize the embedding of the different modal box and diamond operators
(shown here only for {\isasymbox}\isactrlsub a and \isactrlbold {\isasymdiamond}\isactrlsub a).
Each of them is embedded as a function from sentence meanings to
sentence meanings (type $``m{\Rightarrow}m"$), and they employ (restricted)
quantification over possible worlds, following a Kripke
semantics.\footnote{Note that in addition to the ASCII name ``cjboxa'',
	Isabelle/HOL supports graphical
	notation ``{\isacharparenleft}{\isachardoublequoteopen}\isactrlbold
	{\isasymbox}\isactrlsub
	a{\isacharunderscore}{\isachardoublequoteclose}{\isacharparenright}''.
This is essential  for
	obtaining intuitive mathematical representations.}
\small
\\
\isanewline
\isacommand{abbreviation}%
\ cjboxa\ {\isacharcolon}{\isacharcolon}\ {\isachardoublequoteopen}m{\isasymRightarrow}m{\isachardoublequoteclose}\ {\isacharparenleft}{\isachardoublequoteopen}\isactrlbold {\isasymbox}\isactrlsub a{\isacharunderscore}{\isachardoublequoteclose}{\isacharparenright}\ \isakeyword{where}\ {\isachardoublequoteopen}\isactrlbold {\isasymbox}\isactrlsub a{\isasymphi}\ {\isasymequiv}\ {\isasymlambda}c\ w{\isachardot}\ {\isasymforall}v{\isachardot}\ {\isacharparenleft}av\ w{\isacharparenright}\ v\ {\isasymlongrightarrow}\ {\isacharparenleft}{\isasymphi}\ c\ v{\isacharparenright}{\isachardoublequoteclose}\isanewline
\isacommand{abbreviation}%
\ cjdiaa\ {\isacharcolon}{\isacharcolon}\ {\isachardoublequoteopen}m{\isasymRightarrow}m{\isachardoublequoteclose}\ {\isacharparenleft}{\isachardoublequoteopen}\isactrlbold {\isasymdiamond}\isactrlsub a{\isacharunderscore}{\isachardoublequoteclose}{\isacharparenright}\ \isakeyword{where}\ {\isachardoublequoteopen}\isactrlbold {\isasymdiamond}\isactrlsub a{\isasymphi}\ {\isasymequiv}\ {\isasymlambda}c\ w{\isachardot}\ {\isasymexists}v{\isachardot}\ {\isacharparenleft}av\ w{\isacharparenright}\ v\ {\isasymand}\ {\isacharparenleft}{\isasymphi}\ c\ v{\isacharparenright}{\isachardoublequoteclose}
\normalsize
\\
\\
The following definitions correspond to the semantical embedding of DDL deontic operators in Isabelle/HOL.
The first one represents conditional obligations of the form
``\isasymphi\ must be the case given \isasymsigma" and is embedded as a
dyadic relation (type $``m{\Rightarrow}m{\Rightarrow}m"$). The second
and third represent the so-called ``actual" and ``ideal" obligations. 
\small
\\
\isanewline%
\isacommand{abbreviation}%
\ cjod\ {\isacharcolon}{\isacharcolon}\ {\isachardoublequoteopen}m{\isasymRightarrow}m{\isasymRightarrow}m{\isachardoublequoteclose}\ {\isacharparenleft}{\isachardoublequoteopen}\isactrlbold O{\isasymlangle}{\isacharunderscore}{\isacharbar}{\isacharunderscore}{\isasymrangle}{\isachardoublequoteclose}{\isacharparenright}\ \isakeyword{where}\ {\isachardoublequoteopen}\isactrlbold O{\isasymlangle}{\isasymphi}{\isacharbar}{\isasymsigma}{\isasymrangle}\ {\isasymequiv}\ {\isasymlambda}c\ w{\isachardot}\ ob\ {\isacharparenleft}{\isasymsigma}\ c{\isacharparenright}\ {\isacharparenleft}{\isasymphi}\ c{\isacharparenright}{\isachardoublequoteclose}\isanewline
\isacommand{abbreviation}%
\ cjoa\ {\isacharcolon}{\isacharcolon}\ {\isachardoublequoteopen}m{\isasymRightarrow}m{\isachardoublequoteclose}\ {\isacharparenleft}{\isachardoublequoteopen}\isactrlbold O\isactrlsub a{\isacharunderscore}{\isachardoublequoteclose}{\isacharparenright}\ \isakeyword{where}
\isanewline
\ \ \ \ {\isachardoublequoteopen}\isactrlbold O\isactrlsub a{\isasymphi}\ {\isasymequiv}\ {\isasymlambda}c\ w{\isachardot}\ {\isacharparenleft}ob\ {\isacharparenleft}av\ w{\isacharparenright}{\isacharparenright}\ {\isacharparenleft}{\isasymphi}\ c{\isacharparenright}\ {\isasymand}\ {\isacharparenleft}{\isasymexists}x{\isachardot}\ {\isacharparenleft}av\ w{\isacharparenright}\ x\ {\isasymand}\ {\isasymnot}{\isacharparenleft}{\isasymphi}\ c\ x{\isacharparenright}{\isacharparenright}{\isachardoublequoteclose}\isanewline
\isacommand{abbreviation}%
\ cjop\ {\isacharcolon}{\isacharcolon}\ {\isachardoublequoteopen}m{\isasymRightarrow}m{\isachardoublequoteclose}\ {\isacharparenleft}{\isachardoublequoteopen}\isactrlbold O\isactrlsub i{\isacharunderscore}{\isachardoublequoteclose}{\isacharparenright}\ \isakeyword{where}
\isanewline
\ \ \ \ {\isachardoublequoteopen}\isactrlbold O\isactrlsub i{\isasymphi}\ {\isasymequiv}\ {\isasymlambda}c\ w{\isachardot}\ {\isacharparenleft}ob\ {\isacharparenleft}pv\ w{\isacharparenright}{\isacharparenright}\ {\isacharparenleft}{\isasymphi}\ c{\isacharparenright}\ {\isasymand}\ {\isacharparenleft}{\isasymexists}x{\isachardot}\ {\isacharparenleft}pv\ w{\isacharparenright}\ x\ {\isasymand}\ {\isasymnot}{\isacharparenleft}{\isasymphi}\ c\ x{\isacharparenright}{\isacharparenright}{\isachardoublequoteclose}
\normalsize
\subsection{Logical Validity (Classical)}
The SSE technique also allows us to embed different notions of logical
validity: 
context-dependent modal validity and general validity
(modal validity in each context).
\small
\\
\isanewline%
\isacommand{abbreviation}%
\ modvalidctx\ {\isacharcolon}{\isacharcolon}\ {\isachardoublequoteopen}m{\isasymRightarrow}c{\isasymRightarrow}bool{\isachardoublequoteclose}\ {\isacharparenleft}{\isachardoublequoteopen}{\isasymlfloor}{\isacharunderscore}{\isasymrfloor}\isactrlsup M{\isachardoublequoteclose}{\isacharparenright}\ \isakeyword{where}\ {\isachardoublequoteopen}{\isasymlfloor}{\isasymphi}{\isasymrfloor}\isactrlsup M\ {\isasymequiv}\ {\isasymlambda}c{\isachardot}\ {\isasymforall}w{\isachardot}\ {\isasymphi}\ c\ w{\isachardoublequoteclose}\ %
\isanewline
\isacommand{abbreviation}%
\ modvalid\ {\isacharcolon}{\isacharcolon}\ {\isachardoublequoteopen}m{\isasymRightarrow}bool{\isachardoublequoteclose}\ {\isacharparenleft}{\isachardoublequoteopen}{\isasymlfloor}{\isacharunderscore}{\isasymrfloor}{\isachardoublequoteclose}{\isacharparenright}\ \isakeyword{where}\ {\isachardoublequoteopen}{\isasymlfloor}{\isasymphi}{\isasymrfloor}\ {\isasymequiv}\ {\isasymforall}c{\isachardot}\ {\isasymlfloor}{\isasymphi}{\isasymrfloor}\isactrlsup M\ c{\isachardoublequoteclose}
\normalsize
\subsection{Kaplan's Context Features} \label{sectionKaplan}
\noindent{Kaplan's theory, originally named ``Logic of Demonstratives (LD)''
	\citep{Kaplan1979,Kaplan1989-KAPA}, aims at modeling the behavior of certain context-sensitive linguistic expressions like
	the pronouns `I', `my', `it', the demonstrative pronouns `that', `this', the adverbs `here', `now', `tomorrow', the adjectives `actual', `present', and others. Such expressions are known as \textit{indexicals} and so Kaplan's logical system, among others, is usually referred to as a ``logic of indexicals".}
	
It is characteristic of an indexical that its content varies with context, i.e.~they have a context-sensitive character.
Non-indexicals have a fixed character. 
LD models context-sensitivity by representing contexts as
quadruples of features:
\isa{{\isasymlangle}Agent{\isacharparenleft}c{\isacharparenright}{\isacharcomma}\
  Position{\isacharparenleft}c{\isacharparenright}{\isacharcomma}\
  World{\isacharparenleft}c{\isacharparenright}{\isacharcomma}\ Time{\isacharparenleft}c{\isacharparenright}{\isasymrangle}}.
\hskip-.5em\ The agent and the position of context \emph{c} can be seen as the actual speaker
and place of the utterance respectively, while \emph{c}'s world and time stand for the circumstances of evaluation of the	expression's content and allow for the interaction of indexicals with alethic and tense modalities respectively.
To keep things simple, 
we restrict ourselves to representing a context \emph{c} as the pair:
\isa{{\isasymlangle}Agent{\isacharparenleft}c{\isacharparenright}{\isacharcomma}\
	World{\isacharparenleft}c{\isacharparenright}{\isasymrangle}}
      and model the functional concepts ``Agent" and ``World"
as uninterpreted logical constants. An extension of our work to operate on
	Kaplan's context quadruples is straightforward.
\small
\\
\isanewline%
\isacommand{consts}%
\
Agent{\isacharcolon}{\isacharcolon}{\isachardoublequoteopen}c{\isasymRightarrow}e{\isachardoublequoteclose}\
\ %
\isamarkupcmt{function retrieving the agent corresponding to context
	c%
}
\isanewline
\isacommand{consts}%
\
World{\isacharcolon}{\isacharcolon}{\isachardoublequoteopen}c{\isasymRightarrow}w{\isachardoublequoteclose}\
\ %
\isamarkupcmt{function retrieving the world corresponding to context
	c%
}
\normalsize
\subsection{Indexical Validity}
\noindent{Kaplan's notion of (context-dependent) logical truth for a sentence corresponds to its context-sensitive formula (of type ``$m$", i.e.~``$c{\Rightarrow}w{\Rightarrow}bool$") being true in the given context and at its corresponding world.
Kaplan's notion of logical validity for a sentence requires its truth in all contexts. This notion is also known as indexical validity.}%
\small
\\
\isanewline%
\isacommand{abbreviation}%
\ ldtruectx{\isacharcolon}{\isacharcolon}{\isachardoublequoteopen}m{\isasymRightarrow}c{\isasymRightarrow}bool{\isachardoublequoteclose}\ {\isacharparenleft}{\isachardoublequoteopen}{\isasymlfloor}{\isacharunderscore}{\isasymrfloor}\isactrlsub {\isacharunderscore}{\isachardoublequoteclose}{\isacharparenright}\ \isakeyword{where}\ {\isachardoublequoteopen}{\isasymlfloor}{\isasymphi}{\isasymrfloor}\isactrlsub c\ {\isasymequiv}\ {\isasymphi}\ c\ {\isacharparenleft}World\ c{\isacharparenright}{\isachardoublequoteclose}
\isanewline
\isacommand{abbreviation}%
\ ldvalid{\isacharcolon}{\isacharcolon}{\isachardoublequoteopen}m{\isasymRightarrow}bool{\isachardoublequoteclose}\ {\isacharparenleft}{\isachardoublequoteopen}{\isasymlfloor}{\isacharunderscore}{\isasymrfloor}\isactrlsup D{\isachardoublequoteclose}{\isacharparenright}\ \isakeyword{where}\ {\isachardoublequoteopen}{\isasymlfloor}{\isasymphi}{\isasymrfloor}\isactrlsup D\ {\isasymequiv}\ {\isasymforall}c{\isachardot}\ {\isasymlfloor}{\isasymphi}{\isasymrfloor}\isactrlsub c{\isachardoublequoteclose}%
\normalsize
\\
\\
\noindent{The following lemmas show that indexical validity is indeed weaker than its classical modal counterpart (truth at all worlds for all contexts).}%
\small
\\
\isanewline%
\isacommand{lemma}%
\ {\isachardoublequoteopen}{\isasymlfloor}A{\isasymrfloor}\ {\isasymLongrightarrow}\ {\isasymlfloor}A{\isasymrfloor}\isactrlsup D{\isachardoublequoteclose}%
\ %
\isacommand{by}%
\ simp%
\ \isamarkupcmt{proven using Isabelle's term-rewriting engine (simp)}%
\isanewline
\isacommand{lemma}%
\ {\isachardoublequoteopen}{\isasymlfloor}A{\isasymrfloor}\isactrlsup D\ {\isasymLongrightarrow}\ {\isasymlfloor}A{\isasymrfloor}{\isachardoublequoteclose}\ \isacommand{nitpick}%
\ %
\isacommand{oops}%
\ %
\isamarkupcmt{countermodel}%
\normalsize
\\
\\
\noindent{The \emph{countermodel} computed by the model finder \emph{Nitpick}
	\citep{Nitpick} for the latter lemma
	consists of one context $c_1$ and two worlds $w_1$ and $w_2$; where
	World$(c_1)=w_1$ and where $A$ holds for $c_1$  and $w_1$, but not
	for $c_1$  and $w_2$ (\emph{Nitpick} returns further
	insightful details which we omit here). }
\noindent{Below we use \emph{Nitpick}  to show that the interplay between indexical validity and the DDL modal and deontic operators does not result in \emph{modal collapse}. Moreover, we show that the necessitation rule does not work for
	indexical validity (in contrast to classical modal validity as
	defined for DDL).%
\small
\\
\isanewline%
\isacommand{lemma}
\ {\isachardoublequoteopen}{\isasymlfloor}P\isactrlbold {\isasymrightarrow}\isactrlbold O\isactrlsub aP{\isasymrfloor}\isactrlsup D{\isachardoublequoteclose}\ \isacommand{nitpick}
\ %
\isacommand{oops}
\ %
\isamarkupcmt{countermodel for deontic modal collapse found}%
\isanewline
\isacommand{lemma}
\ {\isachardoublequoteopen}{\isasymlfloor}P\ \isactrlbold {\isasymrightarrow}\ \isactrlbold {\isasymbox}\isactrlsub aP{\isasymrfloor}\isactrlsup D{\isachardoublequoteclose}\ \isacommand{nitpick}
\ %
\isacommand{oops}
\ %
\isamarkupcmt{countermodel for alethic modal collapse found}%
\isanewline%
\isacommand{lemma}
{\isachardoublequoteopen}{\isasymlfloor}A{\isasymrfloor}\isactrlsup D\ {\isasymLongrightarrow}\ {\isasymlfloor}\isactrlbold
{\isasymbox}\isactrlsub aA{\isasymrfloor}\isactrlsup
D{\isachardoublequoteclose}\ \isacommand{nitpick}\  \isacommand{oops}
\isamarkupcmt{countermodel for necessitation rule found}
\normalsize
\\
\\
\noindent{Below we introduce a kind of ``a priori necessity'' operator (to be contrasted to the more
	traditional alethic necessity). This operator satisfies the necessitation rule for indexical validity.\footnote{Note that {\isasymbox}\isactrlsup D is not part of Kaplan's
		original system. It has been added by us in order to better
		highlight some semantic features of our formalization of Gewirth's
		theory in the next section and for enabling the use of the
		necessitation rule for drawing inferences.}
	In Kaplan's framework, a sentence being logically (i.e. indexically) valid means its being true \emph{a priori}: It is guaranteed to be true
	in every possible context in which it is uttered, even though it may express distinct propositions (i.e. contents or intensions) in different contexts. This correlation
	between indexical validity and \emph{a prioricity} has also been claimed in other two-dimensional semantic frameworks \citep{SEP2DSem}.}%
\small
\\
\isanewline%
\isacommand{abbreviation}%
\ ldvalidbox\ {\isacharcolon}{\isacharcolon}\ {\isachardoublequoteopen}m{\isasymRightarrow}m{\isachardoublequoteclose}\ {\isacharparenleft}{\isachardoublequoteopen}\isactrlbold {\isasymbox}\isactrlsup D{\isacharunderscore}{\isachardoublequoteclose}{\isacharparenright}\ \isakeyword{where}\ {\isachardoublequoteopen}\isactrlbold {\isasymbox}\isactrlsup D{\isasymphi}\ {\isasymequiv}\ {\isasymlambda}c\ w{\isachardot}\ {\isasymlfloor}{\isasymphi}{\isasymrfloor}\isactrlsup D{\isachardoublequoteclose}
\isanewline
\isacommand{lemma}%
\ NecLD{\isacharcolon}\ {\isachardoublequoteopen}{\isasymlfloor}A{\isasymrfloor}\isactrlsup D\ {\isasymLongrightarrow}\ {\isasymlfloor}\isactrlbold {\isasymbox}\isactrlsup DA{\isasymrfloor}\isactrlsup D{\isachardoublequoteclose}%
\ \ %
\isacommand{by}%
\ simp%
\ %
\isamarkupcmt{necessitation rule proven (term-rewriting)%
}%
\normalsize

\subsection{Quantification}
\noindent{By utilizing Isabelle/HOL's parameterized types (rank-1
	polymorphism), we can easily enrich our logic with (first-order and higher-order) quantifiers.}%
\\
\isanewline
\small
\isacommand{abbreviation}%
\ mforall{\isacharcolon}{\isacharcolon}{\isachardoublequoteopen}{\isacharparenleft}{\isacharprime}t{\isasymRightarrow}m{\isacharparenright}{\isasymRightarrow}m{\isachardoublequoteclose}\ {\isacharparenleft}{\isachardoublequoteopen}\isactrlbold {\isasymforall}{\isachardoublequoteclose}{\isacharparenright}\ \isakeyword{where}\ {\isachardoublequoteopen}\isactrlbold {\isasymforall}{\isasymPhi}\ {\isasymequiv}\ {\isasymlambda}c\ w{\isachardot}{\isasymforall}x{\isachardot}\ {\isacharparenleft}{\isasymPhi}\ x\ c\ w{\isacharparenright}{\isachardoublequoteclose}\isanewline
\isacommand{abbreviation}%
\ mexists{\isacharcolon}{\isacharcolon}{\isachardoublequoteopen}{\isacharparenleft}{\isacharprime}t{\isasymRightarrow}m{\isacharparenright}{\isasymRightarrow}m{\isachardoublequoteclose}\ {\isacharparenleft}{\isachardoublequoteopen}\isactrlbold {\isasymexists}{\isachardoublequoteclose}{\isacharparenright}\ \isakeyword{where}\ {\isachardoublequoteopen}\isactrlbold {\isasymexists}{\isasymPhi}\ {\isasymequiv}\ {\isasymlambda}c\ w{\isachardot}{\isasymexists}x{\isachardot}\ {\isacharparenleft}{\isasymPhi}\ x\ c\ w{\isacharparenright}{\isachardoublequoteclose}
\normalsize
\\
\\
This definition of embedded parametric quantifiers (which reuses
$\lambda$-abstraction to avoid the explicit introduction of a new
binding mechanism) follows earlier
work \citep{J23}. However, it is defined here for Kaplan's
sentence meanings and in this sense constitutes another relevant extension of
previous work.

\section{Representing Gewirth's Ethical Theory}
\label{sec:representation}
\noindent{In this section we encode  and mechanize  Gewirth's (\citeyear{GewirthRM}) ethical theory ---respectively, ethical argument--- which aims at justifying an upper moral principle called the ``Principle of Generic Consistency" (PGC). In a nutshell, according to this principle, any intelligent agent (by virtue of its self-understanding as an agent)
is rationally committed to asserting that (i) it has rights to freedom and well-being,
and (ii) all other agents have those same rights. The argument used by Gewirth to derive the PGC (presented in detail in \cite{GewirthRM,Beyleveld}) is by no means trivial and has stirred much controversy in legal and moral philosophy during the last decades. It has also been discussed in political philosophy as an argument for the \textit{a priori} necessity of human rights \citep{Beyleveld2012}. Perhaps more relevant for us, the PGC has lately been proposed as a means to bound the impact of artificial general intelligence (AGI) by 
\cite{Kornai}.

Kornai draws on Gewirth's PGC as the paradigmatic principle which, assuming it can reliably be represented in a machine, will enable the design of a safety mechanism of a mathematical nature that ensures that an AGI will always respect basic human's rights over all other things.
This is based on the assumption that such an intelligent agent is able to recognize itself, as well as humans, as agents acting voluntarily on self-chosen purposes, i.e.~as what Gewirth calls: prospective purposive agents (PPA). Every agent designed to follow the PGC will thus be deductively committed, on pain of self-contradiction, to acting in accord with the \textit{generic} rights (i.e. to freedom and well-being) of all agents.\footnote{Our work constitutes a most relevant first step for further assessment of Kornai's claim. E.g.~we plan to embody our encoding of Gewirth's theory in virtual agents and devise and conduct respective empirical studies. The merits of the work presented here are however not tied to the validity of Kornai's claim. We illustrate that representation and reasoning with complex ethical theories is meanwhile feasible to an extent unmatched before; and this is highly relevant for implementing explicit ethical intelligent systems. In the following, we will present some commented extracts of our formal encoding of Gewirth's theory and of the computer-supported verification of the argument leading to the PGC.} 

\subsection{Gewirth's Ethical Theory}

Gewirth's meta-ethical position is known as moral (or ethical) rationalism. According to it, moral principles are knowable \emph{a priori}, by reason alone. Immanuel Kant is  the most famous figure who has defended such a position. He argued for the existence of upper moral principles (e.g.~his ``categorical imperative") from which we can reason
in a top-down fashion to deduce and evaluate other more concrete maxims and actions.
In contrast to Kant, Gewirth derives such upper moral principles by starting from purely logical (i.e.~non-moral) considerations alone. The argument for the PGC employs what Gewirth calls ``the dialectically necessary method" within the ``internal viewpoint" of an agent. Although the logical inferences leading to the PGC are drawn relative to the reasoning agent, 
\cite{GewirthRM} further argues that
\emph{\small ``the dialectically necessary method propounds the contents of this relativity as necessary ones, since the statements it presents reflect judgements all agents necessarily make on the basis of what is necessarily
involved in their actions \ldots\ The statements the method attributes to the agent are set forth as necessary ones in that they reflect what is conceptually
necessary to being an agent who voluntarily or freely acts for purposes he wants to attain."}
In other words, the ``dialectical necessity" of the assertions and inferences made in the argument comes from the definitional features (i.e.~conceptual analysis) of the involved notions of agency, purposeful action, obligation, rights, etc. In order to adequately represent this informal notion of \textit{a priori} dialectical/analytic necessity, we resorted to the formal notion of \textit{indexical validity} as developed in David Kaplan's logical framework LD \citep{Kaplan1979,Kaplan1989-KAPA}.
\\
\indent{The cogency of Gewirth's theory will be put to the test in
  Section \ref{sec:reasoning} by using it to reconstruct his argument
  (with minor fixes) for the PGC as logically valid. However, we first
  need to introduce the basic theory itself. To get some inspiration
  we study the main steps of Gewirth's argument (with original numbering from \cite{Beyleveld}):}


\begin{description} \setlength\itemsep{0em}
	\item[(1)] \textbf{[Premise]} I act voluntarily for some (freely chosen) purpose E ---equivalent by definition to: I am a prospective purposive agent (PPA). 
	\item[(2)] E is (subjectively) good ---i.e.~I value E proactively.
	\item[(3)] My freedom and well-being (FWB) are generically necessary conditions of my agency ---i.e.~I need them to achieve any purpose whatsoever.
	\item[(4)] My FWB are necessary goods (at least for me).
	\item[(5)] I have (maybe nobody else does) a claim right to my FWB.
	\item[(13)] \textbf{[Conclusion]} Every PPA has a claim right to their FWB.
\end{description}

In his informal proof, Gewirth claims that the latter generalization step (from ``I" to all agents) is done on purely logical grounds and does not presuppose any kind of universal moral principle, and his result is meant to hold with some kind of necessity.\footnote{We were indeed able to formally verify Gewirth's claim, on condition of committing to an alternative notion of (logical) necessity: Kaplan's ``indexical validity''.}
In this respect, Deryck Beyleveld, author of an authoritative book on Gewirth's theory (\citeyear{Beyleveld}), comments on its first page:
\emph{\small
``[Gewirth's] argument purports to establish the PGC as a rationally necessary proposition with an apodictic status
\emph{for any PPA} equivalent to that enjoyed by the logical principle of noncontradiction itself."} 

\indent{In what follows, we provide some \textit{meaning postulates}\footnote{Definitions and axiomatized conceptual interrelations framing the inferential role of terms. We also refer to them as ``explications". Meaning postulates were introduced in \cite{carnap1952meaning}. }
for the core ethical concepts used to articulate both the PGC and the argument leading to it (as outlined above). We illustrate how to exploit the expressivity of our embedded object logic (DDL enhanced with quantifiers and contexts) to \emph{intuitively} represent and mechanize such a complex ethical theory for the first time in a computer. We also illustrate the utilization of interactive proof assistants (Isabelle/HOL) to assess the argument and to reason with Gewirth's theory.}

\subsection{Agency}
\noindent{Since Isabelle/HOL is a based on a Church's functional type theory, we need to assign all terms a type. We give ``purposes" the same type as sentence meanings (type `$c{\Rightarrow}w{\Rightarrow}bool$' aliased `m'), so that ``acting on a purpose" is
represented analogously to having a certain propositional attitude (like ``desiring that so and so \ldots"). The terms ``ActsOnPurpose" and ``NeedsForPurpose" obtain functional types, and thus expressions like ``(ActsOnPurpose A E)" and ``(NeedsForPurpose A P E)" are read as ``agent A acts on purpose E" and ``agent A needs to have property P in order to reach purpose E".
We also define a type alias $p$ for properties (functions mapping individuals to characters). }
\\
\isanewline
\small
\isacommand{type{\isacharunderscore}synonym}%
\ p\ {\isacharequal}\ {\isachardoublequoteopen}e{\isasymRightarrow}m{\isachardoublequoteclose}\ %
\ %
\isamarkupcmt{function from individuals to sentence meanings (characters)}%
\isanewline
\isacommand{consts}%
\ ActsOnPurpose{\isacharcolon}{\isacharcolon}\ {\isachardoublequoteopen}e{\isasymRightarrow}m{\isasymRightarrow}m{\isachardoublequoteclose}\ %
\isanewline
\isacommand{consts}%
\ NeedsForPurpose{\isacharcolon}{\isacharcolon}\ {\isachardoublequoteopen}e{\isasymRightarrow}p{\isasymRightarrow}m{\isasymRightarrow}m{\isachardoublequoteclose}
\normalsize
\\
\\
\noindent{In Gewirth's argument, an individual with agency (i.e.~capable of purposive action) is said to be a PPA (prospective purposive agent). This definition is supplemented with a meaning postulate stating that being a PPA is an essential (i.e.~identity-constitutive) property of an individual. Quite interestingly, this postulate entails a kind of ability for a PPA to recognize other PPAs.\footnote{Lemma ``recognizeOtherPPA'' below is indeed inferred from axiom ``essentialPPA'' using Isabelle's \textit{blast} tactic (a tableaux prover).} For instance, if some individual holds itself as a PPA (seen from its own perspective/context 'd') then this individual `Agent(d)' is considered a PPA from any other agent's perspective/context `c'.}%
\\
\isanewline
\small
\isacommand{definition}%
\ PPA{\isacharcolon}{\isacharcolon}\ {\isachardoublequoteopen}p{\isachardoublequoteclose}\ \isakeyword{where}\ 
\isamarkupcmt{Definition of PPA}
\isanewline
\isacommand{axiomatization}%
\ \isakeyword{where}\ %
essentialPPA{\isacharcolon}\ {\isachardoublequoteopen}{\isasymlfloor}\isactrlbold {\isasymforall}a{\isachardot}\ PPA\ a\ \isactrlbold {\isasymrightarrow}\ \isactrlbold {\isasymbox}\isactrlsup D{\isacharparenleft}PPA\ a{\isacharparenright}{\isasymrfloor}\isactrlsup D{\isachardoublequoteclose}%
\isanewline
\isacommand{lemma}%
\ recognizeOtherPPA{\isacharcolon}\ %
 {\isachardoublequoteopen}{\isasymforall}c\ d{\isachardot}\ {\isasymlfloor}PPA\ {\isacharparenleft}Agent\ d{\isacharparenright}{\isasymrfloor}\isactrlsub d\ {\isasymlongrightarrow}\ {\isasymlfloor}PPA\ {\isacharparenleft}Agent\ d{\isacharparenright}{\isasymrfloor}\isactrlsub c{\isachardoublequoteclose}%
\isanewline
\ \ \ \ \isacommand{using}%
\ essentialPPA\ \isacommand{by}%
\ blast%
\ \isamarkupcmt{proven using Isabelle blast tactic (tableaux)}
\normalsize

\subsection{Goodness}

\noindent{Gewirth's concept of (subjective) goodness applies to purposes and is relative to some agent.
It is thus modeled as a binary relation relating an individual (of type `e') with a purpose (of type `m').
The axioms below are meaning postulates interrelating the concept of goodness with agency and are given as indexically valid sentences (in Kaplan's sense).\footnote{Their higher-order and modal nature well illustrates the need for expressive knowledge representation and reasoning techniques.}
In particular, we have noticed the need to postulate a further axiom (\textit{explGoodness3}), which represents the intuitive notion of ``seeking the good" by asserting
that, from an agent's perspective, necessarily good purposes are not only action motivating,
but also entail an instrumental obligation to their realization (but only where possible).}%
\\
\isanewline
\small
\isacommand{consts}%
\ Good{\isacharcolon}{\isacharcolon}{\isachardoublequoteopen}e{\isasymRightarrow}m{\isasymRightarrow}m{\isachardoublequoteclose}
\isanewline
\isacommand{axiomatization}\ \isakeyword{where}
\isanewline
\ \ \ \ explGoodness{\isadigit{1}}{\isacharcolon}
\ {\isachardoublequoteopen}{\isasymlfloor}\isactrlbold {\isasymforall}a\ P{\isachardot}\ ActsOnPurpose\ a\ P\ \isactrlbold {\isasymrightarrow}\ Good\ a\ P{\isasymrfloor}\isactrlsup D{\isachardoublequoteclose}
\isanewline
\ \ \ \ explGoodness{\isadigit{2}}{\isacharcolon}
\ {\isachardoublequoteopen}{\isasymlfloor}\isactrlbold {\isasymforall}P\ M\ a{\isachardot}\ Good\ a\ P\ \isactrlbold {\isasymand}\ NeedsForPurpose\ a\ M\ P\ \isactrlbold {\isasymrightarrow}\ Good\ a\ {\isacharparenleft}M\ a{\isacharparenright}{\isasymrfloor}\isactrlsup D{\isachardoublequoteclose}
\isanewline
\ \ \ \ explGoodness{\isadigit{3}}{\isacharcolon}
\ {\isachardoublequoteopen}{\isasymlfloor}\isactrlbold {\isasymforall}{\isasymphi}\ a{\isachardot}\ \isactrlbold {\isasymdiamond}\isactrlsub p{\isasymphi}\ \isactrlbold {\isasymrightarrow}\ \isactrlbold O{\isasymlangle}{\isasymphi}\ {\isacharbar}\ \isactrlbold {\isasymbox}\isactrlsup DGood\ a\ {\isasymphi}{\isasymrangle}{\isasymrfloor}\isactrlsup D{\isachardoublequoteclose}%
\normalsize

\subsection{Freedom and Well-Being}

\noindent{According to Gewirth, enjoying freedom and well-being (which we take together as the predicate ``FWB") is the \emph{contingent} property which represents the ``necessary conditions" or ``generic features" of agency
(i.e.~FWB is \textit{always} required in order to be able to act on \textit{any} purpose whatsoever).
As before, we take this as an \textit{a priori}
characteristic of  FWB and therefore axiomatize it as an indexically valid sentence.
The last two axioms postulate that FWB is a contingent property.}%
\\
\isanewline
\small
\isacommand{consts}%
\ FWB{\isacharcolon}{\isacharcolon}{\isachardoublequoteopen}p{\isachardoublequoteclose}\ %
\isamarkupcmt{FWB is a property (has type \isa{e{\isasymRightarrow}m})}
\isanewline
\isacommand{axiomatization}%
\ \isakeyword{where}
\isanewline
\ \ \ \ explicationFWB{\isadigit{1}}{\isacharcolon}\ {\isachardoublequoteopen}{\isasymlfloor}\isactrlbold {\isasymforall}P\ a{\isachardot}\ NeedsForPurpose\ a\ FWB\ P{\isasymrfloor}\isactrlsup D{\isachardoublequoteclose}
\isanewline
\ \ \ \ explicationFWB{\isadigit{2}}{\isacharcolon}\ {\isachardoublequoteopen}{\isasymlfloor}\isactrlbold {\isasymforall}a{\isachardot}\ \isactrlbold {\isasymdiamond}\isactrlsub p\ FWB\ a{\isasymrfloor}\isactrlsup D{\isachardoublequoteclose}
\isanewline
\ \ \ \ explicationFWB{\isadigit{3}}{\isacharcolon}\ {\isachardoublequoteopen}{\isasymlfloor}\isactrlbold {\isasymforall}a{\isachardot}\ \isactrlbold {\isasymdiamond}\isactrlsub p\ \isactrlbold {\isasymnot}FWB\ a{\isasymrfloor}\isactrlsup D{\isachardoublequoteclose}%
\normalsize
\subsection{Obligation and Interference%
}
\noindent{Kant's Law (``ought implies can") plays an important role in Gewirth's argument.\footnote{This theorem is indeed derivable directly in DDL from the definition of obligations: If \isa{{\isasymphi}} oughts to obtain then \isa{{\isasymphi}} is possible.} We have noticed the need to slightly amend it in order to render the argument as logically valid. The new variant reads as: ``ought implies \textit{ought to} can".
Our variation is indeed closer to Gewirth's (\citeyear[p.~91-95]{GewirthRM})
textual description, that having an obligation to do X implies that
\emph{\small
``I ought (in the same sense and the same criterion) to be free to do X,
that I ought not to be prevented from doing X, that my capacity to do X ought not to be interfered with." }
\\
\isanewline
\small
\isacommand{lemma}%
\ {\isachardoublequoteopen}{\isasymlfloor}\isactrlbold O\isactrlsub i{\isasymphi}\ \isactrlbold {\isasymrightarrow}\ \isactrlbold {\isasymdiamond}\isactrlsub p{\isasymphi}{\isasymrfloor}{\isachardoublequoteclose}%
\ \isacommand{using}%
\ sem{\isacharunderscore}{\isadigit{5}}ab\ \isacommand{by}%
\ simp~\footnote{Here we use Isabelle's \textit{simp} tool to prove that Kant's lemma follows from one of the DDL semantic conditions (not shown here).}
\isanewline
\isacommand{axiomatization}%
\ \isakeyword{where}\ OIOAC{\isacharcolon}\ {\isachardoublequoteopen}{\isasymlfloor}\isactrlbold O\isactrlsub i{\isasymphi}\ \isactrlbold {\isasymrightarrow}\ \isactrlbold O\isactrlsub i{\isacharparenleft}\isactrlbold {\isasymdiamond}\isactrlsub a{\isasymphi}{\isacharparenright}{\isasymrfloor}\isactrlsup D{\isachardoublequoteclose}
\normalsize
\\
\\
\noindent{Concerning the concept of interference, we have noticed the need to presume that the existence of an individual \textit{b} (successfully) interfering
with some state of affairs $\varphi$ implies that $\varphi$ cannot possibly be obtained in any of the actually possible situations (and the other way round). This axiom implies that if someone (successfully) interferes with agent \textit{a} having FWB, then \textit{a} can no longer possibly enjoy its FWB (and the converse).}
\\
\isanewline
\small
\isacommand{consts}%
\ InterferesWith{\isacharcolon}{\isacharcolon}{\isachardoublequoteopen}e{\isasymRightarrow}m{\isasymRightarrow}m{\isachardoublequoteclose}\ %
\isanewline
\isacommand{axiomatization}%
\ \isakeyword{where}\ explicationInterference{\isacharcolon}\ %
{\isachardoublequoteopen}{\isasymlfloor}{\isacharparenleft}\isactrlbold {\isasymexists}b{\isachardot}\ InterferesWith\ b\ {\isasymphi}{\isacharparenright}\ \isactrlbold {\isasymleftrightarrow}\ \isactrlbold {\isasymnot}\isactrlbold {\isasymdiamond}\isactrlsub a{\isasymphi}{\isasymrfloor}{\isachardoublequoteclose}%
\isanewline
\isacommand{lemma}%
\ InterferenceWithFWB{\isacharcolon}\ {\isachardoublequoteopen}{\isasymlfloor}\isactrlbold {\isasymforall}a{\isachardot}\ {\isacharparenleft}\isactrlbold {\isasymexists}b{\isachardot}\ InterferesWith\ b\ {\isacharparenleft}FWB\ a{\isacharparenright}{\isacharparenright}\ \isactrlbold {\isasymleftrightarrow}\ \isactrlbold {\isasymnot}\isactrlbold {\isasymdiamond}\isactrlsub a{\isacharparenleft}FWB\ a{\isacharparenright}{\isasymrfloor}{\isachardoublequoteclose}%
\isanewline
\ \ \ \ \isacommand{using}%
\ explicationInterference\ \isacommand{by}%
\ blast
\normalsize
\subsection{Rights and Other-Directed Obligations%
}
\noindent{Gewirth (\citeyear[p.~66]{GewirthRM}) points out the existence of a correlation between an agent's own claim rights and other-referring obligations.
A claim right is a right which entails duties or obligations for other agents regarding the right-holder
(so-called Hohfeldian claim rights in legal theory).
We model this concept of claim rights in such a way that an individual \textit{a} has a (claim) right to having some property $\varphi$ if and only if it is obligatory that every
(other) individual \textit{b} does not interfere with the state of affairs $(\varphi\ a)$.
Since there is no particular individual to whom this directive is addressed, this obligation has been referred to by Gewirth as being ``other-directed" (aka.~``other-referring") in contrast to ``other-directing" obligations
which entail a moral obligation for some particular subject \cite[p.~41,~51]{Beyleveld}.
This latter distinction is essential to Gewirth's argument.}%
\\
\isanewline
\small
\isacommand{definition}%
\ RightTo{\isacharcolon}{\isacharcolon}{\isachardoublequoteopen}e{\isasymRightarrow}{\isacharparenleft}e{\isasymRightarrow}m{\isacharparenright}{\isasymRightarrow}m{\isachardoublequoteclose}\ \isakeyword{where}\ %
{\isachardoublequoteopen}RightTo\ a\ {\isasymphi}\ {\isasymequiv}\ \isactrlbold O\isactrlsub i{\isacharparenleft}\isactrlbold {\isasymforall}b{\isachardot}\ \isactrlbold {\isasymnot}InterferesWith\ b\ {\isacharparenleft}{\isasymphi}\ a{\isacharparenright}{\isacharparenright}{\isachardoublequoteclose}%
\normalsize
\\
\\
\noindent{Now that all axioms of the theory are in place, we need to show that
they are indeed logically consistent. For this we use Isabelle's model finder \emph{Nitpick} to compute a corresponding model (not shown here) having one context, one individual and two worlds.}%
\\
\isanewline
\small
\isacommand{lemma}%
\ True\ \isacommand{nitpick}%
{\isacharbrackleft}satisfy{\isacharcomma}\ card\ c {\isacharequal} {\isadigit{1}}{\isacharcomma}\ card\ e {\isacharequal} {\isadigit{1}}{\isacharcomma}\ card\ w {\isacharequal} {\isadigit{2}}{\isacharbrackright}%
\ \isacommand{oops}\ \isamarkupcmt{model found}%
\normalsize

\section{Reasoning with Gewirth's Ethical Theory}
\label{sec:reasoning}
\noindent The PGC can be seen as a particular variant (or emendation) of the \textit{golden rule}:  treating others as one's self would wish to be treated. A self-acknowledged agent (i.e.~a PPA) would read the PGC as a moral commandment:
``I ought to act in accord with the generic rights of my recipients as well as of myself" \cite[p.~153]{GewirthRM}.	
Urging a fellow human being to obey such a principle without having explained its deeper rationale will presumably at best elicit an absent-minded, cursory acknowledgment. The difficulty here lies not only in the lack of understanding or agreement of what the given words mean (what is a ``generic right”?), but also in the addressee's lack of `immersion' in the underlying conceptual framework and the inferential practices behind such a principle (an unaware addressee would not be able to infer a third-party obligation from a right claim). In short, any moral principle \textit{qua sentence} makes best sense in the context of the background theory from which it is obtained as a well-founded part; this has been argued e.g. by the philosopher W. V. O. Quine
in his holistic view of meaning (cf. \citeyear{quine2013word}).

This situation is not much different for machines. In order to correctly interpret and apply an ethical principle, we need to (i)
determine the meaning of its constituent concepts (action/agency, right, freedom and well-being, etc.); and (ii) determine the meaning of other relevant concepts (goodness, necessity, interference, obligation, etc.) playing a role in its articulation (and justification) within the underlying theory. Talk of meanings can be obscure, so let us put it in model-theoretical terms: The set of models of the logical theory has to be constrained to properly fit the target conceptualization (i.e.~to only entail intended models). These constraints are set by meaning postulates, i.e.~axioms and definitions. Their adequacy can be assessed by studying the extent to which they enable the validation (or invalidation) of candidate theorems (or non-theorems). As is already known, the main theorem we aim at validating here is the PGC, suitably paraphrased as: \textit{Every PPA has a claim right to its freedom and well-being.} The reconstructed proof in Isabelle/HOL of the theorem below is shown in Fig.~\ref{FigGewirthProof}.
\\
\isanewline
\small
\noindent{\isacommand{theorem}%
\ PGC{\isacharcolon}\ \isakeyword{shows}\ %
{\isachardoublequoteopen}{\isasymforall}C{\isachardot}\ {\isasymlfloor}PPA\ {\isacharparenleft}Agent\ C{\isacharparenright}\ \isactrlbold {\isasymrightarrow}\ {\isacharparenleft}RightTo\ {\isacharparenleft}Agent\ C{\isacharparenright}\ FWB{\isacharparenright}{\isasymrfloor}\isactrlsub C{\isachardoublequoteclose}}
\normalsize
\\
\\
\noindent{In Sections \ref{sec:embedding} and
  \ref{sec:representation}, besides from formally articulating
  Gewirth's theory, we have used some of Isabelle's proof methods
  (simp, blast, etc.)  and the \textit{Nitpick} model finder to verify
  some relevant inferences and to guarantee consistency, thus the
  theory's adequacy has already partly been assessed.  In addition, we
  have used a combination of interactive and automated theorem proving
  to reconstruct Gewirth's argument for the PGC as logically valid by
  formally proving it within the complex logical framework built so
  far.  We thus contribute an exemplary case study illustrating how to
  reason with highly-expressive formal representations of complex,
  natural-language ethical theories by harnessing the power of
  higher-order theorem provers (drawing on the SSE approach).  In the
  argument's reconstruction as displayed in
  Fig.~\ref{FigGewirthProof}, some of the intermediate inference steps
  leading to the main conclusion (PGC) have indeed been hinted at by
  automated tools; cf.~\cite{Sources,GewirthProofAFP} for further details. In particular, some missing implicit premises (not considered in Gewirth's original argument) have been uncovered, namely the explications of the concepts of \textit{goodness} and \textit{interference} and the amendment to Kant's Law: ``ought implies \textit{ought to} can". Note that the mechanized argument matches the granularity-level as can also be found in human constructed informal arguments, and all the sub-arguments (sub-proofs) can automatically be found by automated theorem proving technology. Moreover, the whole proof as presented can be automatically verified using a standard laptop in under a second.

\begin{figure}[th]\centering
	\includegraphics[width=11cm,height=10.6cm]{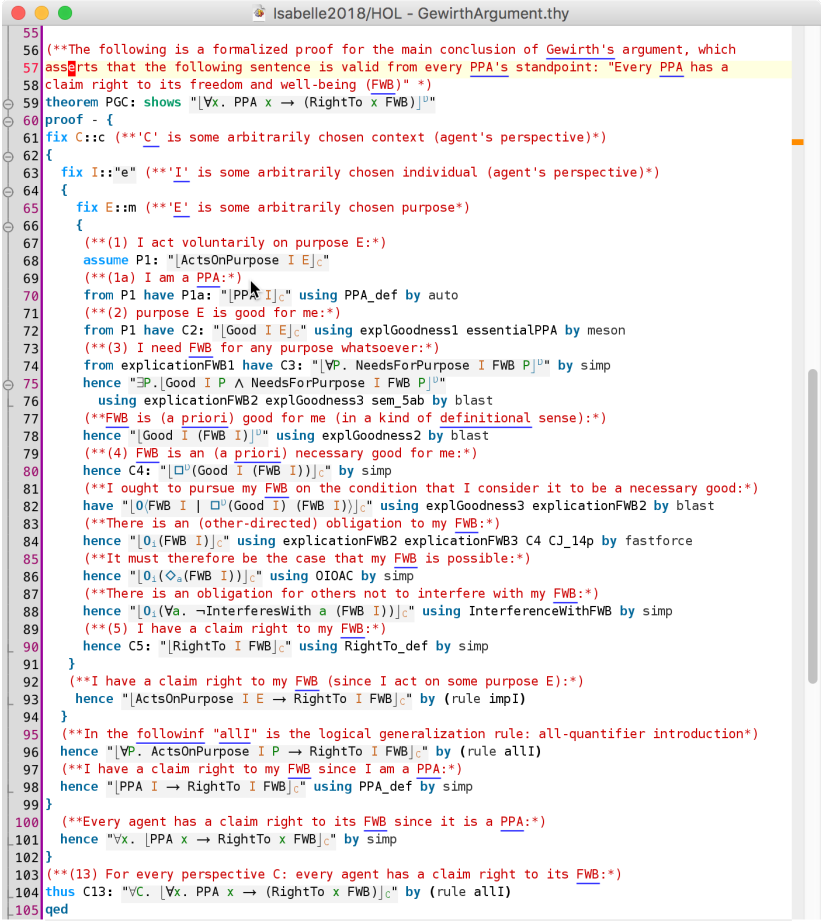}
	\caption{Gewirth's proof encoded in the Isabelle/HOL proof assistant.} \label{FigGewirthProof}
\end{figure}


\section{Related Work and Summary} \label{sec:Summary}
We achieve several improvements over related work such
as 
\cite{bringsjord2006toward} and 
\cite{furbach2015deontic}:
(i) Due the use of enriched DDL (enabled by our hi<gher-order
meta-logic) we are not suffering from contrary-to-duty issues; (ii) we
make use of truly higher-order encodings as required for the
adequate modeling of the PGC; (iii) we overcome unintuitive,
machine-oriented formula representations; and (iv) we do not stop with
supporting proof automation, but combine it with intuitive
user interaction. Combinations of (i)--(iv) also apply to
more recent related work 
by
\cite{GovindarajuluBringsjordIJCAI17}, \cite{hooker2018toward} and
\cite{pereira2016programming}, 
which are not applicable to complex theories like Gewirth's PGC without
considering significant simplifications (accepting
e.g.~contrary-to-duty issues is potentially dangerous).

Utilizing a semantical embedding of a suitable combination of
expressive non-classical logics in meta-logic HOL, an ambitious ethical
theory, Gewirth's PGC, has exemplarily
been encoded and mechanized on
the computer. Our methodology supports both highly intuitive
representation of and interactive-automated reasoning with the encoded
theory. Automated theorem provers have even helped to reveal some hidden issues
in Gewirth's argument.  The presented methodology is motivating research in
different, albeit related, directions: (i) for conducting analogous formal
assessments of further ambitious ethical theories, and (ii) for progressing with the implantation
of explicit ethical reasoning competencies in future intelligent
autonomous systems \emph{by adapting state-of-the-art theorem proving
technology and by combining the expertise of different research communities}.



\bibliographystyle{plainnat}
\sloppy 

\end{document}